\documentclass[twocolumn,prl,showpacs]{revtex4}
\usepackage{epsfig}
\usepackage{graphicx}
\usepackage{dcolumn}
\usepackage{bm}

\begin{document}

\title{Size segregation in granular media induced by phase transition
}
\author{M. Tarzia$^{a}$, A. Fierro$^{a,b}$, M. Nicodemi$^{a,b}$, M. Pica
Ciamarra$^{a}$, and A. Coniglio$^{a,b}$}
\affiliation{${}^a$ Dipartimento di Scienze Fisiche, Universit\`{a} degli Studi
di Napoli ``Federico II'', INFM and INFN, via Cinthia, 80126 Napoli, Italy}

\affiliation{${}^b$ INFM - Coherentia, Napoli, Italy}

\date{\today}

\begin{abstract}
In order to study analytically the nature of the size segregation in granular
mixtures, we introduce a mean field theory in the framework of a
statistical mechanics approach, based on Edwards' original ideas. For simplicity
we apply the theory to a lattice model for hard sphere binary mixture under
gravity, and we find
a new purely thermodynamic mechanism which gives rise to
the size segregation phenomenon.
By varying the number of small grains and the mass ratio, we find a crossover 
from Brasil nut to reverse Brasil nut effect, which becomes a true phase
transition when the number of small grains is larger then a critical value.
We suggest that this transition is induced by the effective attraction between
large grains due to the presence of small ones (depletion force).
Finally the theoretical results are confirmed by numerical simulations of
the $3d$ system under taps.
\end{abstract}

\maketitle

In the last decades a great attention has been devoted to the
study of the problem of vertically shaken granular mixtures under
gravity. It was observed that such systems depending on the
control parameters can mix or segregate their components
spontaneously according to a mechanism which is still largely
unclear, although of deep practical and conceptual relevance
\cite{rev_segr,nagel}. Rosato et al. \cite{rosato} showed that
large grains surrounded by a sea of smaller ones rise to the top
when subjected to vertical shaking. This is the well known
``Brazil Nut Effect'' (BNE), while the opposite one (i.e. large
grains on the bottom and small ones on the top) is known as
``Reverse Brazil Nut Effect'' (RBNE) \cite{luding,breu}. The BNE
was originally explained \cite{rosato} in terms of a geometric
effect, based on  ``percolation'' arguments.
Along with geometry, dynamical effects,
such as convection \cite{knight} or  inertia \cite{shinbrot},
were shown to play a crucial role.
The hydrodynamic equations for binary mixtures \cite{jenkins} were also applied,
and in this framework the regimes of BNE and RBNE \cite{herrmann} were found
to depend on the presence of inelastic dissipation.

Recent results finally outlined that segregation processes may involve
global mechanisms, such as condensation \cite{luding} or
phase separation \cite{mullin}.
This suggested a change of perspective on the issue and
the idea to formulate a statistical mechanics description of
these phenomena \cite{luding,both,seg}.
To this aim, in Ref.s~\cite{luding, both} the interplay between
size and mass was studied and a phase diagram for BNE/RBNE transition
based on the competition between percolation and condensation was proposed.

In the present paper, applying a statistical mechanics approach for
non-thermal systems \cite{Edwards, e1}, a new purely thermodynamic
mechanism is found responsible for size segregation. We follow a
statistical mechanics approach for granular materials under taps,
which was recently developed \cite{nfc} on the basis of Edwards'
original ideas \cite{Edwards, e1}. This approach postulates that
time averages coincide with suitable ensemble averages over the
mechanically stable states (i.e. those where the system is found
at rest). Such hypothesis was shown \cite{nfc} to hold with good
approximation in a lattice model for granular mixtures under taps.
In order to study size segregation, in this framework we apply a mean field 
theory to such model.


We find that, by fixing the number of grains of the two species, it is
possible to obtain a smooth crossover from BNE to RBNE by
changing the mass ratio of the grains. However if the number of
small grains increases above a critical value, a sharp
transition occurs resulting in a phase separation between the two
species: Vertical size segregation is induced by the presence of gravity
which forces the heavier phase to move downward. At the critical
point, critical fluctuations are found typical of a second order
phase transition. We suggest that  this demixing
phase separation is due to the presence of the depletion force.
This is an effective attraction between large grains, obtained
by tracing over all the degrees of freedom of small grains. The
strength of the interaction is proportional to the number of small
grains. The depletion force, typically present in thermal
binary mixtures \cite{Evans}, in fact may be also present in $3d$
granular mixtures under gravity \cite{depletion}. The
theoretical results are confirmed by numerical simulations of the
$3d$ system.

{\em The model~}-- The model \cite{nfc,seg} is a hard sphere binary mixture 
under gravity
made up of two species, $1$ (small) and $2$ (large), respectively with
diameter $a_0=1$ and $\sqrt{2} a_0$, and masses $m_1$ and $m_2$.
In order to simplify the calculations we divide the space in
cubic cells of linear size $a_0$ whose vertices can be or not
occupied by grains. Each site of the lattice is labeled by the two indices
$i=1,\dots,L^2$ and $z=1,\dots,L_z$, where $L$ and $L_z$ are respectively the
horizontal and vertical dimensions of the lattice.
A microstate of the model is completely identified by the occupancy variables
$\{n_i^z\}$, associated to each site of the lattice:
$n_i^z= 0, 1, 2$
respectively if the site $(i,z)$ is empty, filled by a small grain or by a
large one.
In terms of these variables the model Hamiltonian is
\begin{equation}
{\cal H}={\cal H}_{HC}+ m_1gH_1 + m_2gH_2,
\label{H1}
\end{equation}
where $H_1=\sum_{i,z} z \delta_{n_i^z 1}$,
$H_2=\sum_{i,z} z \delta_{n_i^z 2} $ are respectively the heights of the two
species,
and ${\cal H}_{HC}$ is the hard core potential, preventing two nearest
neighbor sites to be both occupied if at least one contains a large grain.

In Ref.s~\cite{nfc} it was shown that the model, Eq.~(\ref{H1}), under taps is
described with good approximation by a statistical mechanics approach based
on Edwards' originally ideas. In particular it was found that
the weight of a microscopic state, $\{n_i^z\}$, is given by
\begin{equation}
P(\{n_i^z\})= \Pi (\{n_i^z\})\cdot e^{-{\cal H}_{HC}(\{n_i^z\})-
\beta_1m_1gH_1 -\beta_2m_2gH_2},
\end{equation}
where $\Pi (\{n_i^z\})$ is a projector over the
``mechanically stable'' states, namely $\Pi (\{n_i^z\})=1$ if $\{n_i^z\}$
is mechanically stable, and $\Pi (\{n_i^z\}) =0$ otherwise,
$\beta_1$ and $\beta_2$ are the variables conjugated respectively to the
gravitational energies of the two species, and
the two ``configurational temperatures'' $T_1=\beta_1^{-1}$ and
$T_2=\beta_2^{-1}$ are increasing functions of the tap amplitude.
The system partition function is given by ${\cal Z}=\sum_{\{n_i^z\}}
P(\{n_i^z\})$
where the sum is over all the microstates but, due to the projector,
only the mechanically stable ones are taken into account.
In the following calculations we adopt a simple definition of ``mechanical
stability'': a grain is considered stable if it has a grain underneath
\cite{nota_proj}.

{\em Mean field approximation~}-- 
In the present paper the partition function, ${\cal Z}$,
is evaluated  in mean field approximation:
we consider a generalization of Bethe-Peierls method for anisotropic systems
(due to gravity) already used in previous papers \cite{seg,epl,jamming}.
In particular, we consider a Bethe lattice made up by
$L_z$ horizontal layers (i.e., $z\in\{1,...,L_z\}$) whose edges can be
occupied by grains. Each layer is a random graph of given
connectivity, $k-1=4$. Each site in layer $z$ is also
connected to its homologous site in $z-1$ and $z+1$
(the total connectivity is thus $k+1=6$).
We use the grancanonical ensemble by introducing two chemical potentials,
$\mu_1$ and $\mu_2$, relative to species $1$ and $2$.

In the following we give the main ideas of the calculations and
refer to a longer paper \cite{jamming} for the details (see also
\cite{MP}). The Bethe-Peierls recursion equations, which allow to calculate
the partition function, are written in terms of
the local ``cavity fields'' defined by:
$e^{s_{n}^{(i,z)}} = Z_{n,s}^{(i,z)}/Z_{0,s}^{(i,z)}$,
$e^{u_{n}^{(i,z)}} = Z_{n,u}^{(i,z)}/Z_{0,u}^{(i,z)}$,
$e^{v_{n}^{(i,z)}} = \overline{Z}_{n,u}^{(i,z)}/Z_{0,u}^{(i,z)}$,
$e^{d_{n}^{(i,z)}} = Z_{n,d}^{(i,z)}/Z_{0,d}^{(i,z)}$,
$e^{c_{n}^{(i,z)}} = \overline{Z}_{n,d}^{(i,z)}/Z_{0,d}^{(i,z)}$
(with $n=1,2$).
Here $Z_{0,s}^{(i,z)}$ and $Z_{n,s}^{(i,z)}$ are the partition functions
of the ``side'' branch restricted respectively to configurations
where the site $(i,z)$ is empty or filled by a particle of specie $n$;
analogously, $Z_{n,u}^{(i,z)}$, $Z_{0,u}^{(i,z)}$ and
$\overline{Z}_{n,u}^{(i,z)}$ (resp. $Z_{n,d}^{(i,z)}$, $Z_{0,d}^{(i,z)}$
and $\overline{Z}_{n,d}^{(i,z)}$) are the partition functions of the
``up'' (resp. ``down'') branch restricted to configurations where
the site $(i,z)$ is filled by a grain of species $n$, empty with the upper
(resp. lower) site empty and empty with the upper (resp. lower) site
filled by a grain of species $n$.

The calculations are limited to the ``fluid'' phase corresponding
to a solution of Bethe-Peierls equations
where local fields in each layer are site independent \cite{nota_cryst}.
Such a solution, characterized by horizontal translational invariance, is
given by the fixed points of the following equations:
\begin{eqnarray} \label{ricorrenza}
&&\!\!\!\!\!\!\!\!\!\!\!\!e^{s_{n}^{(z)}}  =
\phi_n^{(z)}
\left[
{\cal K}_n^{(z)}
\right]^{k-2}
\frac{P_n^{(z+1)} R_n^{(z-1)}}{Q^{(z)}},
\\
\nonumber
&&\!\!\!\!\!\!\!\!\!\!\!\!e^{u_{n}^{(z)}} =
\phi_n^{(z)}
\left[
{\cal K}_n^{(z)}
\right]^{k-1} P_n^{(z+1)},
\,\,\,\,\,\,\,\,
e^{d_{n}^{(z)}}  = e^{u_{n}^{(z)}} \frac{R_n^{(z-1)}}{P_n^{(z-1)}},
\\
\nonumber
&&\!\!\!\!\!\!\!\!\!\!\!\!e^{v_{n}^{(z)}}=e^{u_{n}^{(z+1)}},
\qquad \,\,\,\,\,\,\,
e^{c_{n}^{(z)}}  =  e^{d_{n}^{(z-1)}} \left[1 + R_2^{(z-1)}
\right]^{-1},
\end{eqnarray}
where $n=1,2$, and $\phi_n^{(z)} = e^{\beta_n(\mu_n-m_ngz)}$.
The functions appearing in Eq.~(\ref{ricorrenza}) have a simple expression
in terms of the local fields \cite{nota_func}.
From the local fields we calculate the free energy \cite{nota_free}, $F$, and
all the quantities of interest, such as
density profiles of small and large grains, $\sigma_1 (z)$ and $\sigma_2 (z)$,
the grain number per unit surface, $N_{n} = \sum_z \sigma_n (z)$, and their
average height,
$h_n=\langle{z_n}\rangle=\sum_z z \sigma_n(z) / \sum_z \sigma_n(z)$
(with $n=1,2$).
\begin{figure}[ht]
\begin{center}
\epsfig{figure=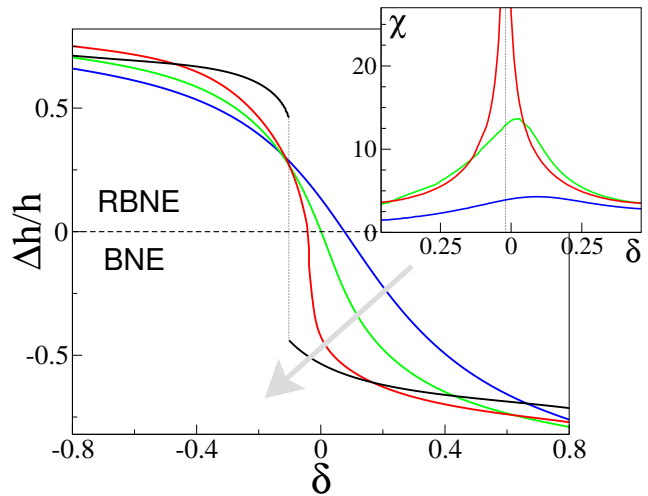,scale=0.3,angle=270}
\end{center}
\vspace{-0.3cm}
\caption{(color online). {\bf Main frame} 
$\frac{\Delta h}{h}$ as a function of $\delta$ in
the mean field approximation for fixed value of $N_2=~0.685$, and $N_1=~0.5$ 
(blue curve), $N_1=~0.9$ (green curve), $N_1=~N_{1c}=~1.37$ (red curve) 
and $N_1=~2.4$ 
(black curve) ($N_1$ increases along the arrow). 
{\bf Inset:} The susceptibility, $\chi$, as a function of $\delta$ for fixed 
value of $N_2=~0.685$, and
$N_1=~0.5$ (blue curve), $N_1=~0.9$ (red curve) and $N_1=~N_{1c}=~1.37$ 
(black curve)  ($N_1$ increases from bottom to top).}
\label{fig:deh}
\end{figure}

In the following instead of using the chemical potential variables, $\mu_1$
and $\mu_2$, we use the conjugate variables, $N_1$ and $N_2$, respectively
the number of the small and large grains per unit surface. For simplicity we
fix the configurational temperatures, $T_1=~T_2=~1$, 
and the number of large grains, $N_2$, and vary $N_1$.
In order to study the BNE/RBNE transition we also vary the masses of the 
two species, $m_1$ and $m_2$ (keeping constant $m_1+m_2$), and measure
the height difference, $ \frac{\Delta h}{h} \equiv \frac{h_1 - h_2}{h_1 + h_2}$:
This quantity, that is thus a function of $N_1$ and $\delta \equiv \frac{2 m_1 
- m_2}{2 m_1 + m_2}$, is positive when small grains are in average above 
large grains (RBNE) and negative otherwise (BNE).

\begin{figure}[ht]
\begin{center}
\epsfig{figure=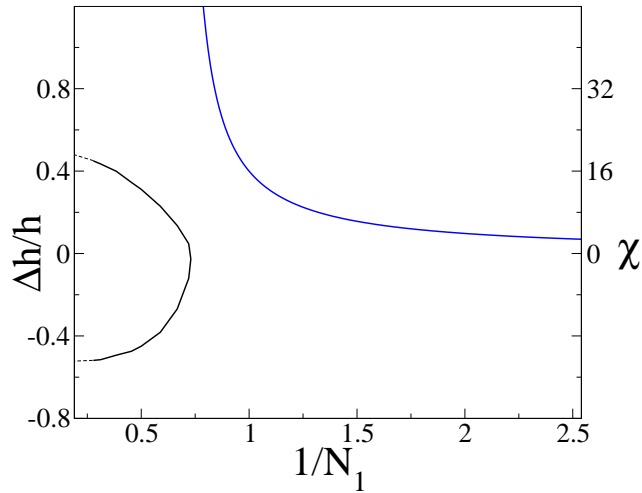,scale=0.3,angle=270}
\end{center}
\vspace{-0.3cm}
\caption{(color online).
The coexistence curve (black curve on the left) and the maximum of the 
susceptibility (blue curve on the right) as a function of $N_1^{-1}$.} 
\label{fig:susc} 
\end{figure}
In the Main Frame of Fig.~\ref{fig:deh},
$ \frac{\Delta h}{h}(\delta,N_1)$ is plotted for fixed $N_1$ as a function
of $\delta$: The different curves correspond to increasing values of $N_1$
(along the arrow). 
We find that for low enough values of $N_1$,
$\frac{\Delta h}{h}$
is a continuous monotonous decreasing function of $\delta$, due to entropic
and buoyancy effect.
By increasing $N_1$ above some critical value ($N_{1c} \simeq
1.37$) $\frac{\Delta h}{h}$ instead exhibits first order transition with a
finite jump from a positive value to a negative one at 
$\delta=\delta_{coex}(N_1)$, where a BNE phase and a RBNE one
coexist with the same free energy. As shown in Fig.~\ref{fig:susc},
the jump becomes smaller and smaller  approaching 
$N_{1c}$ from above, and goes to zero continuously at $N_{1c}$ (the critical 
point is $(\delta_c, N_{1c})$, where $\delta_c\equiv \lim_{N_1\rightarrow 
N_{1c}} \delta_{coex}(N_1)$).

Along the coexistence curve near the critical point in both phases
the order parameter $\frac{\Delta h}{h}\simeq (N_{1c}^{-1}-N_1^{-1})^{\beta}$,
where, solving the mean field equations numerically, $\beta$ is found
consistent with the mean field value $\beta=~0.5$.

As in any second order phase transition, the fluctuations of the order 
parameter diverge at the critical point.
In the Inset of Fig.~\ref{fig:deh}, 
the susceptibility, $\chi(\delta,N_1)\equiv N_{tot}(\langle\Delta h
^2\rangle-\langle\Delta h\rangle^2)$ (where $N_{tot}$ is the total number of
grains) is plotted as a function
of $\delta$ (the curves, from bottom to top, correspond to increasing values 
of $N_1$).  As $N_1$ approaches $N_{1c}$,
the susceptibility diverges at $\delta = \delta_c$.
For $N_1<N_{1c}$, $\chi(\delta,N_1)$ displays a maximum (for the value 
of $\delta$ where $\Delta h$ is zero), which diverges as a power at $N_{1c}$,
$(N_1^{-1}-N_{1c}^{-1})^{-\gamma}$ with an exponent $\gamma$ found, solving
numerically the mean field equations, consistent with
the mean field value $\gamma=~1.0$ (see Fig.~\ref{fig:susc}).

In the present paper we have chosen as free parameters  $N_1$ and
$\delta$: $\delta$ plays the role of an external field, and $N_1$ plays the 
role of an inverse temperature, since it controls the amplitude of the
depletion force \cite{Evans}. To confirm this idea, we have studied the case of
mixture of grains with the same 
radii ($R_2/R_1 = 1$), where the depletion force is completely 
absent. We find that the critical point disappears, and a smooth crossover from
positive to negative values of $\Delta h$, varying the control parameters $N_1$
and $\delta$. 

\begin{figure}[ht]
\begin{center}
\epsfig{figure=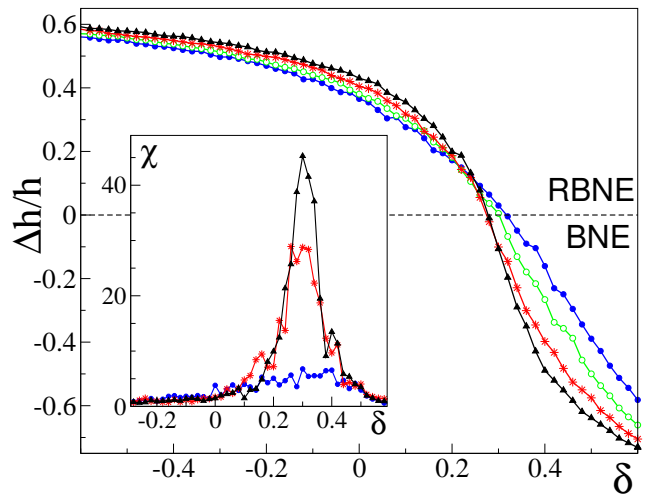,scale=0.3,angle=270}
\end{center}
\vspace{-0.3cm}
\caption{(color online). {\bf Main frame} 
$\frac{\Delta h}{h}$ as a function of $\delta$ in the Monte
Carlo simulations on the cubic lattice for fixed value of $N_2=~1.2$, and
$N_1=~0.6$ (blue curve and filled circles), $N_1=~0.8$ (green curve and empty 
circles), $N_1=~1.0$ (red curve and stars) and $N_1=~1.2$ (black curve and 
triangles). {\bf Inset} The susceptibility, $\chi$, as a function of $\delta$ 
for fixed value of $N_2=~1.2$, and $N_1=~0.6$ (blue curve and filled circles),
$N_1=~1.0$ (red curve and stars) and $N_1=~1.2$ (black curve and
triangles)}
\label{fig:bne}
\end{figure}

{\em Monte Carlo simulations~}-- 
The theoretical results obtained in mean field approximation 
are confirmed by Monte Carlo simulations of the model,
Eq.~(\ref{H1}), under taps \cite{nfc,nota_tap} on a cubic lattice.
In Fig.~\ref{fig:bne}, the height difference, $\frac{\Delta h}{h}$, and the 
susceptibility, $\chi$, are plotted as function of 
$\delta$ for fixed $N_2$ and increasing values of $N_1$.

As found in the mean field approximation, at fixed value of $N_1$, 
$\frac{\Delta h}{h}$ as a function of $\delta$ displays a steeper and steeper 
slope (and the susceptibility, $\chi$, a larger and larger maximum) 
as $N_1$ is increased.
A power law divergence of the maximum of the susceptibility is found at
$N_{1c}\simeq 2.2$, with an exponent $\gamma\simeq 1.6$. 
As usually in mean field
approximation, the critical point ($N_{1c}^{-1}$) is over-estimated.  

As $N_1$ is further increased, $\frac{\Delta h}{h}$ 
obtained by varying $\delta$ at a given rate, becomes irreversible, and a 
strong hysteresis appears, signaling the presence of a first order transition.

{\em Conclusions}-- We focus on the problem of the vertical size
segregation in binary granular mixtures. In the framework of a
statistical mechanics approach to granular media, based on
Edwards' original ideas, we apply a mean field theory to a simple
hard sphere lattice model under gravity, and explain BNE and RBNE
with purely thermodynamic argument. 


We find that by varying the control parameters a transition from BNE to RBNE
can occur as a smooth crossover or as a sharp transition, depending whether
one is above or below the critical point.
The presence of such a critical point is manifested by the divergence
of the height fluctuations of the two species.
Therefore we suggest that it might be important to measure in
experiments these fluctuations in the vicinity of the transition
from BNE to RBNE. The larger the fluctuations the closer to the critical
point are the control parameters.

We would like to thank Hans Herrmann
for many interesting discussions and suggestions.
Work supported by EU Network Number  MRTN-CT-2003-504712, MIUR-PRIN 2004,
MIUR-FIRB 2001, CRdC-AMRA, INFM-PCI.


\begin{thebibliography}{40}
\bibitem{nota_proj} For a given grain configuration, $\{n_i^z\}$,
$\Pi (\{n_i^z\})$ has a tractable expression:
$\Pi (\{n_i^z\}) =\lim_{K\rightarrow\infty}\exp\left\{-K{\cal H}_{Edw}\right\}$
where ${\cal H}_{Edw}=
\sum_{i,z} \left[ \delta_{n_i^z 2}\delta_{n_i^{z-1} 0}\delta_{n_i^{z-2} 0}
+ \delta_{n_i^z 1}\delta_{n_i^{z-1} 0}
\left(1 - \delta_{n_i^{z-2} 2}\right) \right]$.
\bibitem{nota_cryst} Along with the fluid solution, also a crystalline solution
characterized by the breakdown of the horizontal translational invariance,
can be found. Here we consider the case where the crystallization is avoided.
\bibitem{nota_func}
${\cal K}_n^{(z)} = A_n^{(z)} / S^{(z)}$ where
$A_n^{(z)} = 1+\delta_{n,1}e^{s_{1}^{(z)}}$ and
$S^{(z)}=A_1^{(z)}+e^{s_{2}^{(z)}}$. Moreover we have that
$P_n^{(z)} = 1+ \delta_{n,1} e^{u_{1}^{(z)}} + \delta_{n,2} e^{v_{1}^{(z)}}+
e^{v_{2}^{(z)}}$,
$R_1^{(z)} = \delta_{n,1} e^{d_{1}^{(z)}} + \delta_{n,2}e^{c_{1}^{(z)}}
+ e^{c_{2}^{(z)}}$,
$Q^{(z)} = 1+e^{d_{1}^{(z-1)}} 
[1+ e^{u_{2}^{(z+1)}}]
+e^{d_{2}^{(z-1)}} 
[1+e^{u_{1}^{(z+1)}} +  e^{u_{2}^{(z+1)}}]
+ R_2^{(z)}$.
\bibitem{nota_free}
The free energy, $F$, is given by
$F=\sum_{z=0}^H \Delta F_s^{(z)}-\frac{(k-1)}{2}\sum_{z=0}^H
\Delta F_{l,2}^{(z)}-\sum_{z=0}^{H-1}\Delta F_{l,1}^{(z)}$,
where,
$e^{- \Delta F_{s}^{(z)}}=
(S^{(z)}
)^{k-1}Q^{(z)}+\sum_{n=1,2}
\phi_n^{(z)}
( A_n^{(z)}
)^{k-1}P_n^{(z+1)}R_n^{(z-1)}$,
$e^{-\Delta F_{l,1}^{(z)}}=1+\sum_{n=1,2}e^{c_n^{(z)}}
(1+\delta_{n,2}
e^{u_1^{(z+1)}}+e^{u_2^{(z+1)}}
)+\sum_{n=1,2}e^{d_n^{(z)}}P_n^{(z+1)}$,
$e^{-\Delta F_{l,2}^{(z)}}=1+2 e^{s_1^{(z)}}+2 e^{s_2^{(z)}}+e^{2 s_1^{(z)}}$.
\bibitem{nota_tap} Each
tap consists in raising the bath temperature to a finite value
and, after a lapse of time quenching the bath temperature back to zero.
The measurements are performed when the shake is off and the system is at rest.
\end{thebibliography}
\end{document}